\documentclass[jkps,preprint,fleqn,showpacs,showkeys]{revtex4}
\usepackage{graphicx}
\usepackage{subfigure}
\usepackage{amsmath,amssymb}
\usepackage{bm}

\begin{document}
\setcounter{page}{0}

\title[]{One-loop Radiative Corrections to the $\rho$ Parameter in the Left Right Twin Higgs Model}
\author{Jae Yong \surname{Lee}}
\email{littlehiggs@kias.re.kr}
\affiliation{Department of Physics, Korea University, Seoul 136-701, Korea }

\author{ Dong-Won \surname{Jung}}
\email{dwjung@kias.re.kr}
\affiliation{ School of Physics, KIAS, Seoul 130-722, Republic of Korea }

\date[]{}

\begin{abstract}
We implement a one-loop analysis of the $\rho$ parameter in the
Left Right Twin Higgs model, including the logarithmically
enhanced contributions from both heavy fermion and scalar loops.
Numerical analysis indicates that the one-loop corrections are
dominant over the tree-level contributions in most regions of
parameter space. The experimentally allowed values of the
$\rho$-parameter divide the allowed parameter space into two
regions; less than $670~{\rm GeV}$ and larger than $1100~{\rm
GeV}$ roughly, for the symmetry breaking scale $f$.
Therefore our result significantly reduces the parameter space
which are favorably accessible to the LHC.
\end{abstract}


\maketitle

\section{Introduction}
The Standard Model (SM) has excellently described high energy physics up to energies
of ${\mathcal O}(100)$ GeV. The only undetected constituent of the SM to date is 
a Higgs boson which is essential to generating fermion and gauge boson masses.
Theoretically the Higgs boson mass squared is
quadratically sensitive to any new physics scale beyond the Standard Model (BSM) 
and stabilization of the Higgs mass squared prefers the energy scale below ${\mathcal O}(1)$ TeV.
But electroweak precision measurements with naive naturalness assumption
raise the energy scale of the BSM up to 100 TeV or even higher.
Hence, there remains a tension between theory and experiment associated
with the stabilization of the SM Higgs mass. 
With the latest data from the LHC the tension gets stronger.

The basic idea of little Higgs is that the SM Higgs is a pseudo-Nambu-Goldstone
boson(pNGB)~\cite{Georgi:1974yw,Kaplan:1983fs,
Arkani-Hamed:2001nc,Arkani-Hamed:2002pa,Arkani-Hamed:2002qx,
Arkani-Hamed:2002qy,Schmaltz:2002wx}. Stabilization of the Higgs
mass in the little Higgs theories is achieved by the ``collective symmetry breaking"
which naturally renders the SM Higgs mass much smaller than the global symmetry breaking scale.
The distinct elements of little Higgs models are a vector-like heavy top quark, and various
scalar and vector bosons. The former is universal while the latter is model-dependent.
Both of them contribute significantly to one-loop processes and hence set strict constraints on the
parameter space of little Higgs models.
At worst, electroweak precision tests push up the symmetry breaking scale to $5$ TeV or higher,
and even revive the fine-tuning problem of the Higgs potential.

The basic idea of twin Higgs is the same as that of little Higgs.
But the stabilization of the Higgs mass squared is different between the two.
The twin Higgs mechanism introduces additional discrete symmetry
to render no quadratic divergence in the Higgs mass squared.
For instance, the mirror twin Higgs model containing
a complete copy of the SM identifies the discrete symmetry with mirror parity.
The SM world and its mirror world communicate only through the Higgs
so that the mirror particles are very elusive in the SM world 
and yield poor phenomenology at the LHC.

The twin Higgs mechanism can be also realized by identifying the
discrete symmetry with left-right symmetry in the left-right
model~\cite{Chacko:2005un}. The left-right twin Higgs (LRTH) model
contains $U(4)_1\times U(4)_2$ global symmetry as well as
$SU(2)_L\times SU(2)_R\times U(1)_{B-L}$ gauge symmetry. The
left-right symmetry acts on only the two $SU(2)$'s gauge symmetry.
A pair of vector-like heavy top quarks play a key role in
triggering electroweak symmetry breaking just as that of the
little Higgs theories. On top of that, the non-SM Higgs particles acquire
large masses not only at quantum level but also at tree level, making 
the model deliver much richer phenomenology at the LHC~\cite{Goh:2006wj}
compared with the mirror twin Higgs model.
Moreover they lead to large radiative corrections to one-loop processes
so the allowed parameter space can be significantly reduced.
In this paper we perform a one-loop analysis of the $\rho$-parameter in
the LRTH model to reduce the parameter space.

The paper is organized as follows.
The LRTH model is briefly reviewed in Section 2.
The renormalization procedure for $\rho$-parameter is explained in Section 3.
The numerical analysis on the $\rho$-parameter is performed in Section 4.
We present our conclusions in Section 5.
The technical details on the computation of the $\rho$-parameter are reckoned in Appendices.

\section{Left right twin Higgs model in a nutshell}
We review the LRTH model in Ref.~\cite{Goh:2006wj}.
The LRTH model is based on the global $U(4)_1\times U(4)_2$ symmetry,
with a locally gauged subgroup $SU(2)_L\times SU(2)_R\times U(1)_{B-L}$.
A pair of Higgs fields, $H$ and $\hat H$, are introduced and each transforms as
$(4,1)$ and $(1,4)$, respectively, under the global symmetry.
They are written as
\begin{equation}
H=\left(\begin{array}{c} H_L\\ H_R\end{array}\right),\qquad
\hat H=\left(\begin{array}{c} \hat H_L\\ \hat H_R\end{array}\right),
\end{equation}
where $H_{L,R}$ and $\hat H_{L,R}$ are two component objects which are charged under
the $SU(2)_L\times SU(2)_R\times U(1)_{B-L}$ as
\begin{equation}
H_L\mbox{ and }\hat H_L\,:\,(2,1,1),\qquad
H_R\mbox{ and }\hat H_R\,:\,(1,2,1).
\end{equation}
The global $U(4)_1\,(U(4)_2)$ symmetry is spontaneously broken down to its subgroup
$U(3)_1\,(U(3)_2)$ with VEVs
\begin{equation}
\langle H\rangle^T =(0,0,0,f),\qquad
\langle \hat H \rangle^T =(0,0,0,\hat f).
\end{equation}
The spontaneous symmetry breaking results in seven Nambu-Goldstone bosons (NGB),
which are parameterized as
\begin{equation}
H=fe^{\pi/f}\left(\begin{array}{c} 0 \\ 0\\ 0\\ 1\end{array}\right),\qquad\qquad
\pi=\left(\begin{array}{cccc} -\frac{N}{2\sqrt{3}} & 0 & 0 & h_1 \\
0 & -\frac{N}{2\sqrt{3}} & 0 & h_2 \\
0 & 0 & -\frac{N}{2\sqrt{3}} & C \\
h^\ast_1 & h^\ast_2 & C^\ast & \frac{\sqrt{3}N}{2}\end{array}\right),
\end{equation}
where $\pi$ is the corresponding Goldstone field matrix. $N$ is a neutral real field
\footnote{The normalization is naturally altered when applying the unitary gauge.},
$C$ and $C^\ast$ are a pair of charged complex scalar fields, and $h_{SM}=(h_1,h_2)$
is the SM $SU(2)_L$ Higgs doublet.
$\hat H$ is parameterized in the identical way by its own Goldstone boson matrix,
$\hat \pi$, which contains $\hat N$, $\hat C$, and $\hat h=(\hat h^+_1,\hat h^0_2)$.
The two $U(4)/U(3)$'s symmetry breakings yield fourteen NGBs in all.

The linear combination of $C$ and $\hat C$, and the linear combination
of $N$ and $\hat N$ are eaten by the gauge bosons of $SU(2)_R\times U(1)_{B-L}$,
which is broken down to the $U(1)_Y$.
The orthogonal linear combinations, a charged complex scalar $\phi^\pm$
and a neutral real pseudoscalar $\phi^0$, remain as NGBs.
On top of that, the SM Higgs acquires a VEV, $\langle h_{SM}\rangle=(0,v/\sqrt{2})$,
so electroweak symmetry $SU(2)_L\times U(1)_Y$ is broken
down to $U(1)_{EM}$.
But $\hat h$'s do not get a VEV and remain as NGBs.
At the end of the day, the two Higgs VEVs are given by
\begin{equation}\label{eq:VEVs}
\langle H\rangle=\left(\begin{array}{c} 0 \\ if\sin x \\ 0 \\ f\cos x \end{array}\right),\qquad
\langle \hat H\rangle=\left(\begin{array}{c} 0 \\ 0 \\ 0 \\ \hat f \end{array}\right),\
\end{equation}
where $x=\frac{v}{\sqrt{2}f}$. The values of $f$ and $\hat f$ will be bounded by electroweak precision
measurements. In addition, $f$ and $\hat f$ are interconnected
once we set $v=246$ GeV.

\subsection{Gauge sector}
The whole gauge symmetry of the model is
 $SU(3)_C\times SU(2)_L\times SU(2)_R \times U(1)_{B-L}$.
But since $SU(3)_C$ is irrelevant to eletroweak symmetry breaking
$SU(3)_C$ gauge symmetry is not taken into account in this paper.
The generators of the $SU(2)_L\times SU(2)_R\times U(1)_{B-L}$ is
given respectively as,
\begin{equation}
\left(\begin{array}{cc} \frac{1}{2}\sigma_i & 0 \\ 0 & 0\end{array}\right),\qquad
\left(\begin{array}{cc} 0 & 0 \\ 0 & \frac{1}{2}\sigma_i\end{array}\right),\qquad
\frac{1}{2}\left(\begin{array}{cc} 1_2 & 0 \\ 0 & 1_2 \end{array}\right),
\end{equation}
and the corresponding gauge fields are $W^{\pm,0}_L,W^{\pm,0}$ and $B$,
respectively.
The covariant derivative is then given as
\begin{equation}
D_\mu=\partial_\mu-ig{\mathcal W}_\mu-ig' q_{B-L} {\mathcal B}_\mu,
\end{equation}
where
\begin{equation}
{\mathcal W}=\frac{1}{2}\left(\begin{array}{cccc} W^0_L & \sqrt{2}W^+_L & 0 & 0 \\
\sqrt{2}W^-_L & -W^0_L & 0 & 0 \\ 0 & 0 & W^0_R & \sqrt{2}W^+_R\\
0 & 0 & \sqrt{2}W^-_R & -W^0_R \end{array}\right),\qquad
{\mathcal B}=\frac{1}{2}\left(\begin{array}{cccc} B & 0 & 0 & 0 \\ 0& B& 0 & 0 \\
0 & 0 & B & 0 \\ 0 & 0 & 0 & B\end{array}\right),
\end{equation}
and $g$ and $g'$ are the gauge couplings for $SU(2)_{L,R}$ and $U(1)_{B-L}$,
and $q_{B-L}$ is the charge of the field under $U(1)_{B-L}$.

The kinetic term for the two Higgs fields can be written as
\begin{equation}
{\mathcal L}_H=(D_\mu H)^\dagger D^\mu H+(D_\mu \hat H)^\dagger D^\mu \hat H
\end{equation}
with $q_{B-L}=1$.
The Higgs mechanism for both $H$ and $\hat H$ makes
the six gauge bosons massive but one gauge boson, photon, massless.
For the charged gauge bosons, there is no mixing between the $W^\pm_L$ and $W^\pm_R$:
$W^\pm_L$ is identified with the SM weak gauge boson $W^\pm$ while $W^\pm_R$
is much heavier than $W^\pm$ and is denoted as $W^\pm_H$.
Their masses are
\begin{equation}\label{eq:chmass}
M^2_W=\frac{1}{2}g^2f^2\sin^2x,\qquad
M^2_{W_H}=\frac{1}{2}g^2(\hat f^2+f^2\cos^2x).
\end{equation}
Note that $M^2_W+M^2_{W_H}=\frac{g^2}{2}(f^2+\hat f^2)$.
The linear combinations of the neutral gauge bosons $W^0_L, W^0_R$ and $B$
yield two neutral massive gauge bosons $Z, Z_H$ and one photon $A$ with masses, respectively:
\begin{align}
M^2_A&=0,\\
M^2_Z&= \frac{g^2+2g'^2}{g^2+g'^2}\frac{2M^2_WM^2_{W_H}}
{M^2_{W_H}+M^2_W+\sqrt{(M^2_{W_H}-M^2_W)^2+\frac{4g'^2}{g^2+g'^2}M^2_{W_H}M^2_W}},\\
M^2_{Z_H}&= \frac{g^2+g'^2}{g^2}(M^2_W+M^2_{W_H})-M^2_Z.
\end{align}
For later use, we define the Weinberg angle of the LRTH model:
\begin{align}
s_w&=\sin \theta_w=\frac{g'}{\sqrt{g^2+2g'^2}},\\
c_w&=\cos\theta_w=\sqrt{\frac{g^2+g'^2}{g^2+2g'^2}},\\
c2_w&=\sqrt{\cos2\theta_w}=\frac{g}{\sqrt{g^2+2g'^2}}.
\end{align}
The unit of the electric charge is then given by
\begin{equation}
e=gs_w=\frac{gg'}{\sqrt{g^2+2g'^2}}.
\end{equation}

\subsection{Fermion sector}
To cancel the quadratic sensitivity of the Higgs mass to the top quark loops,
a pair of vector-like, charge $2/3$ fermion $({\mathcal Q}_L,{\mathcal Q}_R)$
are incorporated into the top Yukawa sector,
\begin{equation}
{\mathcal L}_{Yuk}=y_L\bar Q_{L3}\tau_2 H^\ast_L {\mathcal Q}_R+
y_R\bar Q_{R3}\tau_2 H^\ast_R {\mathcal Q}_L-M\bar{\mathcal Q}_L {\mathcal Q_R}+h.c.,
\end{equation}
where $\tau_2=\left(\begin{array}{cc} 0 & -1 \\ 1 & 0
\end{array}\right)$, $Q_{L3}=-i(u_{L3},d_{L3})$ and
$Q_{R3}=(u_{R3},d_{R3})$ are the third generation up- and
down-type quarks, respectively. The left-right parity indicates
$y_L=y_R(\equiv y)$. The mass parameter $M$ is essential to the
top mixing. The value of $M$ is constrained by the $Z\to b\bar b$
branching ratio. It can be also constrained by the oblique
parameters, which we will do in the letter. Furthermore, it yields
large log divergence of the SM Higgs mass. To compensate for it
the non-SM gauge bosons also get large masses by increasing the
value of $\hat f$. Therefore it is natural for us to take
$M\lesssim yf$.

Expanding the $H_{L,R}$ field in terms of NGB fields, we acquire
the mass matrix of the fermions. By diagonalizing it we obtain not only the mass
eigenstates for the SM-like and heavy top quarks, but also the mixing angles for the left-handed
and right-handed fermions:
\begin{align}\label{eq:mixtop}
m^2_t&=\frac{1}{2}(M^2+y^2f^2-N_t),\quad
m^2_T=\frac{1}{2}(M^2+y^2f^2+N_t),\\
\sin\alpha_L&=\frac{1}{\sqrt{2}}\sqrt{1-(y^2f^2\cos2x+M^2)/N_t},\\
\sin \alpha_R&=\frac{1}{\sqrt{2}}\sqrt{1-(y^2f^2\cos2x-M^2)/N_t},
\end{align}
where $N_t=\sqrt{(y^2f^2+M^2)^2-y^4f^4\sin^22x}$.
\subsection{Higgs sector}
Among the fourteen NGBs in both $\pi$ and $\hat
\pi$, six NGBs are eaten by the gauge bosons.
The remaining eight NGBs get masses through
quantum effects and/or soft symmetry breaking terms, so called
``$\mu$-term". The Coleman-Weinberg potential, obtained by
integrating out the heavy gauge bosons and top quarks, yields the SM
Higgs potential, which determines the SM Higgs VEV and its mass,
as well as the masses for the other Higgs, $\phi^\pm,\phi^0,\hat h^\pm_1$ and $\hat h^0_2$.
Moreover, the $\mu$-term contributes to the Higgs masses at tree level~\footnote{In the potential, $-\mu_l^2 (H^\dagger_L \hat
H_L + c.c.)$ is possible, but we choose $\mu_l = 0$ for not
spoiling the original motivation of the model and preserving the
stability of $\hat h_2$ dark matter~\cite{Goh:2006wj}.}:
\begin{equation}
V_{\mu}=-\mu^2_r(H^\dagger_R\hat H_R+c.c.)+\hat \mu^2 \hat H^\dagger_L \hat H_L.
\end{equation}

We write down the masses for the Higgs.
\begin{align}
M^2_{\phi^0}&=\frac{\mu^2_rf\hat f}{\hat f^2+f^2\cos^2x}
\Big[\frac{\hat f^2(\cos x+\frac{\sin x}{x}(4+x^2))}{f^2(\cos x+2\frac{\sin x}{x})^2}+
\frac{2\cos x(\cos x+4\frac{\sin x}{x})}{3(\cos x+2\frac{\sin x}{x})}\nonumber\\
&\qquad\qquad\qquad+\frac{f^2\cos^2x(4+\cos x)}{9\hat f^2}\Big],\\
M^2_{\phi^\pm}&=\frac{3}{16\pi^2}\frac{g'^2M^2_{W_H}}{M^2_{Z_H}-M^2_Z}
\Big[\Big(\frac{M^2_W}{M^2_{Z_H}}-1\Big)
{\mathcal Z}(M_{Z_H})-\Big(\frac{M^2_W}{M^2_Z}-1\Big){\mathcal Z}(M_Z)\Big]\nonumber\\
&\,+\frac{\mu^2_rf\hat f}{\hat f^2+f^2\cos^2x}
\Big[\frac{\hat f^2 x^2}{f^2\sin^2x}+2\cos x+\frac{f^2 \cos^3x}{\hat f^2}\Big],\\
M^2_{\hat h_2}&=\frac{3}{16\pi^2}\Big[\frac{g^2}{2}
({\mathcal Z}(M_W)-{\mathcal Z}(M_{W_H}))
+\frac{2g'^2+g^2}{4}\frac{M^2_{W_H}-M^2_W}{M^2_{Z_H}-M^2_Z}
({\mathcal Z}(M_Z)-{\mathcal Z}(M_{Z_H}))\Big]\nonumber\\
&\,+\mu^2_r\frac{f}{\hat f}\cos x+\hat \mu^2,\\
M^2_{\hat h_1}&=M^2_{\hat h_2}
+\frac{3}{16\pi^2}\frac{g'^2M^2_W}{M^2_{Z_H}-M^2_Z}\Big[\Big(\frac{M^2_{W_H}}{M^2_{Z_H}}-1\Big)
{\mathcal Z}(M_{Z_H})-\Big(\frac{M^2_{W_H}}{M^2_Z}-1\Big)
{\mathcal Z}(M_Z)\Big].
\end{align}
where
\begin{equation}
{\mathcal Z}(x)=-x^2(\ln\frac{\Lambda^2}{x^2}+1),
\end{equation}
with $\Lambda$ being a UV cutoff.
The SM Higgs potential arises mainly from both top and
gauge sector. The contribution of fermion loop to the SM Higgs
mass squared is negative and its dominance over the contribution
of gauge boson loops and tree level mass parameter $\mu_r^2
\frac{\hat f}{2 f}$ triggers electroweak symmetry breaking.
We fix the SM Higgs VEV, $v=246~{\rm GeV}$.

\section{The renormalization procedure}
We follow the renormalization procedure in Ref.~\cite{Chen:2003fm} to calculate
the $\rho$-parameter at one-loop order.
The $Z$-pole, $W$-mass, and neutral current data can be used to search for and set limits
on deviations from the SM.
The the $\rho$-parameter is defined as
\begin{equation}\label{eq:defrho}
\rho\equiv\frac{M^2_W}{M^2_Zc^2_\theta}.
\end{equation}

The electroweak mixing angle $s^2_\theta(\equiv 1-c^2_\theta)$ in the effective leptonic (electronic) vertex of the Z boson
is defined as
\begin{equation}\label{eq:sin2theta}
s^2_\theta\equiv \frac{1}{4}\Big(1+\mbox{Re}\,\frac{g^e_V}{g^e_A}\Big),
\end{equation}
in terms of the effective vector and axial vector couplings
$g^e_{V,A}$ of the $Z$ to electrons;
\begin{equation}
{\cal L} = i\bar e \gamma_\mu(g_V+g_A \gamma_5)eZ^\mu.
\end{equation}
The effective Lagrangian of the charged current interaction in the LRTH model
is given by
\begin{equation}
{\mathcal L}_{cc}=\frac{g}{\sqrt{2}}\big(W_{\mu L}^+J^{\mu -}_L+W^-_{\mu L}J^{\mu +}_L\big) + (L\to R),
\end{equation}
where $J^{\mu\pm}_{L,R}$ is the charged currents.
For momenta quite small compared to $M_W$, this effective Lagrangian
gives rise to the effective four-fermion interaction with the Fermi coupling constant,
\begin{equation}
\frac{G_F}{\sqrt{2}}=\frac{g^2}{8M^2_W},
\end{equation}
and the vector and axial vector parts of the neutral current $Zee$
coupling constants are given to the order $v^2/\hat f^2$ as,
\begin{align}
g^e_V&=
\frac{g}{2c_w}\Big[\big(-\frac{1}{2}+2s^2_w\big)+\frac{v^2}{4(f^2+\hat f^2)}
\frac{s^2_w(c2^2_w-2)}{c^4_w}\Big],\\
g^e_A&=
\frac{g}{2c_w}\Big[\frac{1}{2}+\frac{v^2}{4(f^2+\hat f^2)}
\frac{s^2_wc2^2_w}{c^4_w}\Big].
\end{align}
The effective leptonic mixing angle $s^2_\theta$ in Eq.~(\ref{eq:sin2theta})
is then related to the mixing angle $s^2_w$ as
\begin{equation}
s^2_\theta=s^2_w-\frac{v^2}{f^2+\hat f^2}\frac{s^4_w}{c^2_w}.
\end{equation}
It can then be inverted and gives
\begin{equation}
s^2_w=s^2_\theta+\Delta s^2_\theta,
\end{equation}
where
\begin{equation}
\frac{\Delta s^2_\theta}{s^2_\theta}=
-\zeta+\frac{1}{2}\frac{c^2_\theta}{s^2_\theta}
-\sqrt{-\zeta+\Big(\zeta-\frac{1}{2}\frac{c^2_\theta}{s^2_\theta}\Big)^2},
\end{equation}
with
\begin{equation}
\zeta\equiv\frac{v^2}{f^2+\hat f^2}.
\end{equation}

The SM $SU(2)_L$ gauge coupling constant, $g$, is expressed by both the effective leptonic mixing angle, $s^2_\theta$, 
and the fine-structure constant, $\alpha$, as
\begin{equation}
g^2=\frac{e^2}{s^2_w}=\frac{4\pi\alpha}{s^2_\theta}\big(1-\frac{\Delta s^2_\theta}{s^2_\theta}\big).
\end{equation}
The $\rho$-parameter at tree level is
\begin{equation}
\rho^{tree}=\frac{\pi\alpha}{\sqrt{2}G_F c^2_\theta s^2_\theta M^2_Z}
\big(1-\frac{\Delta s^2_\theta}{s^2_\theta}\big).
\end{equation}
The $\rho$-parameter at the tree level is differentiated from unity
and its deviation from unity is of order $v^2/\hat f^2$.

Since the loop factor arising from radiative corrections,
$1/16\pi^2$, is similar in magnitude to $v^2/\hat f^2$ (for $\hat
f\gtrsim 5$ TeV), the one-loop radiative corrections can be
comparable in size to the next-to-leading order corrections at tree level.
At one-loop order the mass relation reads~\cite{Sirlin:1980nh}
\begin{equation}\label{eq:s2c2}
s^2_\theta c^2_\theta=
\frac{\pi\alpha(M_Z)}{\sqrt{2}G_F M^2_Z\rho}\Big[1-\frac{\Delta s^2_\theta}{s^2_\theta}
+\Delta r_Z\Big],
\end{equation}
where $\Delta r_Z$ includes radiative effects from various sources:
\begin{equation} \label{dr}
\Delta r_Z=\frac{\delta \alpha}{\alpha}-\frac{\delta G_F}{G_F}-\frac{\delta M^2_Z}{M^2_Z}
-\Big(\frac{c^2_\theta-s^2_\theta}{c^2_\theta}\Big)\frac{\delta s^2_\theta}{s^2_\theta}.
\end{equation}
We should mention that $\Delta r_Z$ in Eq.~(\ref{dr}) differs from the
usual $\Delta \hat r_Z$ defined in the SM by an extra
corrections due to the renormalization of $s^2_\theta$. In
general, The vertex and box contributions to the radiative effects
are relatively small compared to the other
corrections~\cite{Sirlin:1980nh,Blank:1997qa} and hence we
consider only the so-called ``oblique" type, i.e. the $W$-, $Z$-
and $\gamma$-propagators. The correction due to the vacuum
polarization of the photon, $\delta \alpha$, is given by
\begin{equation}
\frac{\delta \alpha}{\alpha}=\Pi^{\gamma\gamma'}(0)+2\Big(\frac{g^e_V-g^e_A}{Q_e}\Big)
\frac{\Pi^{\gamma Z}(0)}{M^2_Z}.
\end{equation}
Since we ignore the vertex and box corrections, the electroweak radiative correction
to the Fermi constant, $\delta G_F$, stems from the $W$-boson vacuum polarization,
\begin{equation}
\frac{\delta G_F}{G_F}=-\frac{\Pi^{WW}(0)}{M^2_W}.
\end{equation}
The counterterms for the $Z$-boson mass, $\delta M^2_Z$, and the
leptonic mixing angle, $\delta s^2_\theta$, are given by,
respectively~\cite{Blank:1997qa}
\begin{align}
\delta M^2_Z&=Re[\Pi^{ZZ}(M^2_Z)], \\
\frac{\delta s^2_\theta}{s^2_\theta}&=
Re\Big[\Big(\frac{c_\theta}{s_\theta}\Big)\Big\{\frac{\Pi^{\gamma Z}(M^2_Z)}{M^2_Z}
+\frac{v^2_e-a^2_e}{a_e}\Sigma^e_A(m^2_e)
-\frac{v_e}{2s_\theta c_\theta}\Big(\frac{\Lambda^{Z\bar ee}_V(M^2_Z)}{v_e}
-\frac{\Lambda^{Z\bar ee}_A(M^2_Z)}{a_e}\Big)\Big\}\Big],
\end{align}
where $\Lambda^{Z\bar ee}_{V,A}$ are the vector and axial vector form factors of the
unnormalized one-loop $Zee$ vertex corrections, and $\Sigma^e_A$
is the axial part of the electron self-energy.
Once again, we ignore these ``non-oblique" type corrections.

The tree level and radiative corrections except for the W-boson are summed up
and expressed as
\begin{equation}
\Delta \hat r=-\frac{\Delta s^2_\theta}{s^2_\theta}-\frac{Re[\Pi^{ZZ}(M^2_Z)]}{M^2_Z}
+\Pi^{\gamma\gamma'}(0)+2\Big(\frac{g^e_V-g^e_A}{Q_e}\Big)\frac{\Pi^{\gamma Z}(0)}{M^2_Z}
-\frac{c^2_\theta-s^2_\theta}{c_\theta s_\theta}\frac{Re[\Pi^{\gamma Z}(M^2_Z)]}{M^2_Z},
\end{equation}
and then Eq.~(\ref{eq:s2c2}) is rewritten as,
\begin{equation}\label{eq:onedrive}
s^2_\theta c^2_\theta=\frac{\pi\alpha(M_Z)}{\sqrt{2}G_FM^2_Z\rho}\Big[1+\frac{\Pi^{WW}(0)}{M^2_W}
+\Delta \hat r\Big].
\end{equation}
Solving Eq.~(\ref{eq:defrho}) and~(\ref{eq:onedrive})
for the $W$-boson mass we get it as
\begin{equation}\label{eq:mw}
M^2_W=\frac{1}{2}\Big[a(1+\Delta \hat r)+\sqrt{a^2(1+\Delta \hat r)^2+4a\Pi^{WW}(0)}\Big],
\end{equation}
with $ \quad a\equiv\frac{\pi\alpha(M_Z)}{\sqrt{2}G_Fs^2_\theta}$.
Finally the $\rho$ parameter is calculated using Eq.~(\ref{eq:defrho}).

\section{Numerical Analysis}
In order to take the precision measurements,
we need the standard experimental values as input parameters.
These are the input parameters we take~\cite{Yao:2006px}:
\begin{align}
G_F&= 1.16637(1)\times 10^{-5}\mbox{ GeV}^{-2},\\
M_Z&= 91.1876(21)\mbox{ GeV},\\
\alpha(M_Z)^{-1}&=127.918(18),\\
s^2_\theta &= 0.23153(16).
\end{align}
We also take the top and bottom quark masses as
~\cite{Yao:2006px,Rodrigo:1997gy}
\begin{equation}
m_t = 172.3 ~{\rm GeV}, \qquad m_b = 3 ~{\rm GeV},
\end{equation}
where $m_t$ is the central value of the electroweak fit and $m_b$
is the running mass at the $M_Z$ scale with $\overline{MS}$ scheme.
The $\rho$-parameter itself is measured very accurately~\cite{Yao:2006px},
\begin{align}
\rho &\equiv\rho_0\hat \rho\equiv \frac{M^2_W}{M^2_Z c^2_\theta}\\
\rho_0 &= 1.0002^{+0.0007}_{-0.0004}\\
\hat \rho &=1.01043\pm0.00034
\end{align}
Including all the SM corrections (top quark loop, bosonic loops),
 we take the allowed range of $\rho$ parameter as
\begin{equation}
1.00989 ~\leq ~\rho^{exp}~\leq ~1.01026.
\end{equation}
\begin{figure}[htb!]
\centering
\includegraphics[width=4in]{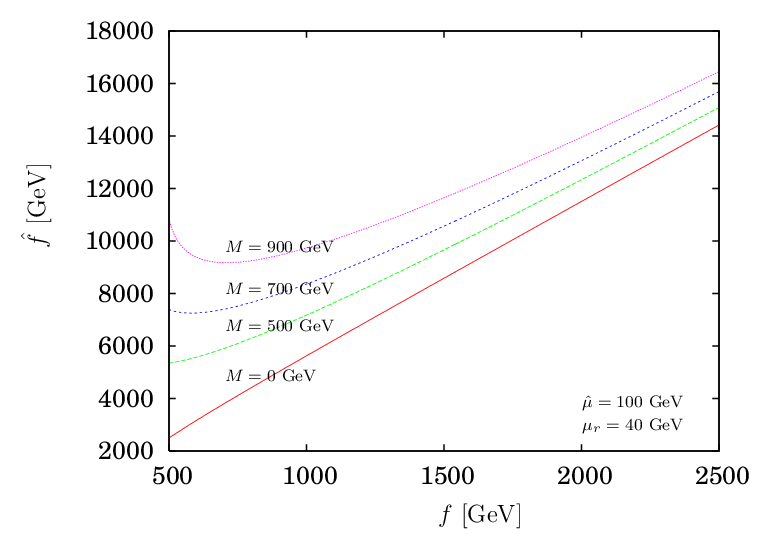}
\caption{\label{fig:ffh}Plots of $\hat f$ versus $f$ with different values of $M$.}
\end{figure}

The input parameters of the LRTH model are as follows;
\begin{equation}
f, ~M, ~\mu_r, ~\hat \mu,
\end{equation}
where $M$ is the heavy top quark mass scale, both $\mu_r$ and
$\hat \mu$ are soft symmetry breaking terms. The masses of the top
and heavy top quarks are determined mainly by $f$ and $M$ while those of
the scalar particles $\hat h_1,~\hat h_2,~\phi^\pm$ and $\phi^0$
largely depend on $\hat \mu,~\mu_r$ and $f$. Another scale $\hat
f$, associated with the masses of the heavy gauge bosons,
can be determined by the electroweak symmetry breaking
condition: there is a generic relation between $\hat f$ and $f$
since Coleman-Weinberg potential of the Higgs boson mostly depends
on $M, f$ and $\hat f$. For scalar potential, there is a tree
level mass term proportional to $\mu_r^2$. So we may not acquire
negative mass squared term which is necessary for electroweak
symmetry breaking and it gives an upper bound for the value of
$\mu_r$.

Figure~\ref{fig:ffh} shows $f$ versus $\hat f$ with various values of the heavy top mass scale, $M$.
For a given $f$, $\hat f$ becomes larger as $M$ increases. 
The heavy top loop through $M$ contributes positively to the Higgs mass 
while the heavy gauge boson loop through $\hat f$ contributes negatively to the Higgs mass.
The two contributions cancel out in order to retain $v=246$ GeV. 
There is also a contribution from tree level mass term $~ \mu_r^2$, but
in most cases it make little difference to the relation as long as $\mu_r$ is much smaller than
the Higgs mass scale. This insensitiveness can be figured out
with simple evaluation. First, from the electroweak symmetry
breaking condition, the mass squared contribution from the soft
symmetry breaking term $\mu_r^2 \frac{\hat f}{2f}$ should be
smaller than that from the fermion loop. This can be written down
approximately as follows:
\begin{equation}
\mu_r^2 \frac{\hat f}{2f} < \frac{3}{8 \pi^2} ( M^2 +y^2 f^2).
\end{equation}
In the above inequality, we ignore the gauge boson loop
contributions since they are small compared to the fermion loop
contributions. In general $\hat f$ is larger than $f$ about $5$
times or more and $\frac{3}{8 \pi^2}$ is very small, we can see
that $\mu_r$ should be very much smaller than $f$. To get the
$\hat f$ which reproduces the electroweak symmetry breaking scale
$v=246$ GeV, we should solve the equation,
\begin{equation}
\frac{3g^4}{64\pi^2} f {\hat f}^2 + \mu_r^2 \hat f +2 \lambda v^2
f -\frac{3}{4\pi^2}f(M^2+f^2)+\frac{3g^4}{64\pi^2} = 0,
\end{equation}
for given $f, M$ and $\mu_r$. $\lambda$ in the above equation is
the coefficient of quartic term and less than $1$ in general. Note
that we derive the above equation with some degree of
approximation. For example, we ignore the logarithmic terms. But
the crude behavior will be similar. In this equation, the
coefficients of $\hat f^2$ and $\hat f$ are much smaller than
constant term, so the solution $\hat f$ is almost insensitive to
the value of $\mu_r$~\footnote{If we rewrite the equation as $a
\hat f^2 + b \hat f +c=0$, the inequality $a, b \ll c$ is
satisfied. In this case the solution is $\hat f \simeq
\sqrt{\frac{c}{a}} -\frac{b}{a} \simeq \sqrt{\frac{c}{a}}$.}.

\begin{figure}[htb!]
\centering
\includegraphics[width=4 in]{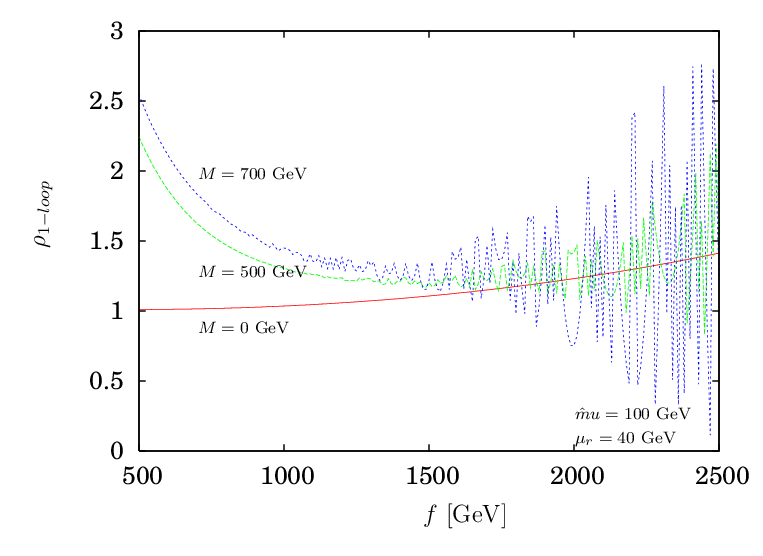}
\caption{The plot for the $\rho$-parameter versus $f$ with different values of $M$.\label{fig:frhos}}
\end{figure}

Plots in Figure~\ref{fig:frhos} illustrate the behavior of one-loop
$\rho$-parameter for various values of $M$.
At $M=0$, where there is no mixing between the top and heavy
top quarks, $\Delta \rho$ increases monotonically with $f$.
For nonzero $M$ where the mixing is turned on, mass of the heavy
top quark becomes large as $M$ increases for a given $f$, and the
fermionic loop contributions tend to become large, too. 
The effects of mixing angles on the fermionic loops
become significant as either $f$ or $M$ increases while the
condition of electroweak symmetry breaking is retained. In other
words, since the mixing angles are determined by $f$ and $M$, the
one-loop corrected $\rho$ parameter begins to waver as $f$
increases even with the fixed scalar mass parameters. Because of
this nature, fine tuning in the $\rho$-parameter is inevitable for
large $f$, as will be shown later.

\begin{figure}[htb!]
\centering
\subfigure[]{
\includegraphics[width=3 in]{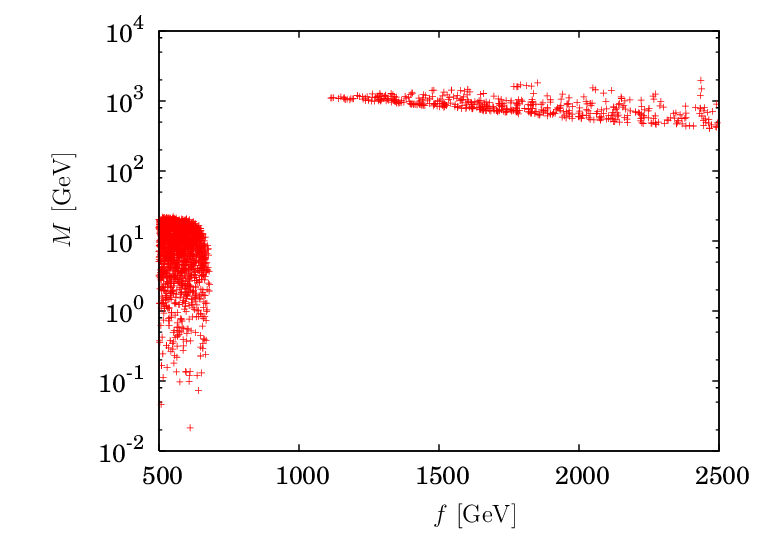}}
\subfigure[]{
\includegraphics[width=3 in]{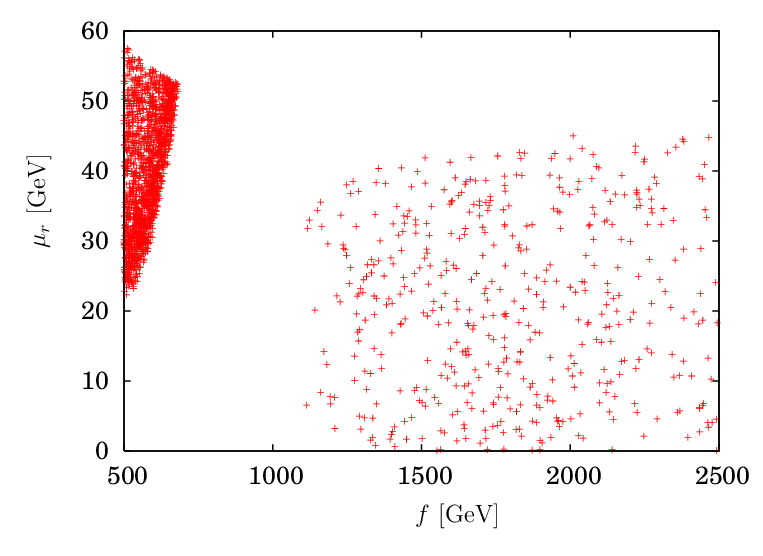}}
\caption{\label{fig:murm} (a) The scatter plot for the mass parameter $M$
with the horizontal axis being the scale parameter $f$.
(b) The scatter plot for the mass parameter $\mu_r$
with the horizontal axis being the scale parameter $f$.}
\end{figure}

To extract a meaningful information on the model parameters from the
$\rho$-parameter, we scan the parameter space generally, i.e.,
\begin{equation}
500 ~{\rm GeV} ~\leq ~f ~\leq ~2500 ~{\rm GeV},~~~~~~ 0~ \leq ~M,
~\mu_r, ~\hat \mu ~\leq ~f.
\end{equation}
Even though too large $f$ makes the model unviable, we take the
rather large value of $f$, 2.5  TeV, as an upper limit for
completeness of the scanning. As a result of $\rho$-parameter
calculation, we can obtain the allowed regions of parameter space.
As an example, Figure~\ref{fig:murm}  shows the allowed regions of
parameter space (a) for $f$ versus $M$ and (b) for $f$ versus
$\mu_r$. It is interesting to notice that the allowed parameter
space is divided into two regions; less than 670 GeV and larger
than 1100 GeV roughly, for $f$. This can be figured out as
follows. The loop corrections tend to be larger as $f$ increases.
It is because the masses of the particles involved in one-loop
correction increase in general as $f$ increases. But at the same
time, the mixing angles of top-heavy top quarks also vary. Since
the mixing angles depend on not only $f$ but also $M$, these two
effects compete during the increase of $f$. Because of this
interplay of top mixing angles and masses, we have two distinct
allowed parameter spaces. For small $f$, solution points prefer
very small values of $M$. It means there is no large mixing
between the top and heavy top quarks. In general, $\Pi^{WW}(0)$ is
large for small $f$, and decreases as $f$ increases.
So for fitting the observed W-boson mass in the small $f$ region,
which is directly related to the $\rho$-parameter,
we restrict the $\Delta \hat r$ within rather
small range. Because the $\Delta \hat r$ is mostly determined by
$\Pi^{ZZ}(M_Z^2)$, it should be also small. For doing that, we
should take the small value of $M$, which makes the masses and
mixing angles of heavy top quark small. We find that in the small
$f$ region, $M$ should be smaller than about $22~{\rm GeV}$.
 Soft symmetry breaking parameter $\mu_r$ is restricted to the
 values less than around $60~{\rm GeV}$.
 This bound arises mainly from the electroweak symmetry breaking
condition, and is generically independent of the $\rho$-parameter.
Another free parameters $\hat \mu$ is not restricted from the
one-loop corrected $\rho$-parameter. The reason is that $\hat \mu$
only contributes to the masses of $\hat h_1$ and $\hat h_2$, and
their contributions are effectively cancelled among the relevant
loop diagrams. It is pointed out in Appendix C.

\begin{figure}[htb!]
\centering 
\subfigure[]{\includegraphics[width=2 in]{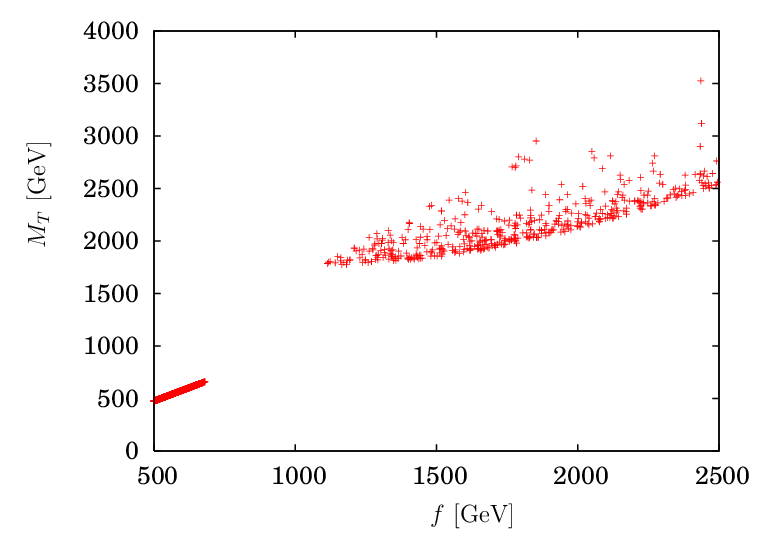}} 
\subfigure[]{\includegraphics[width=2 in]{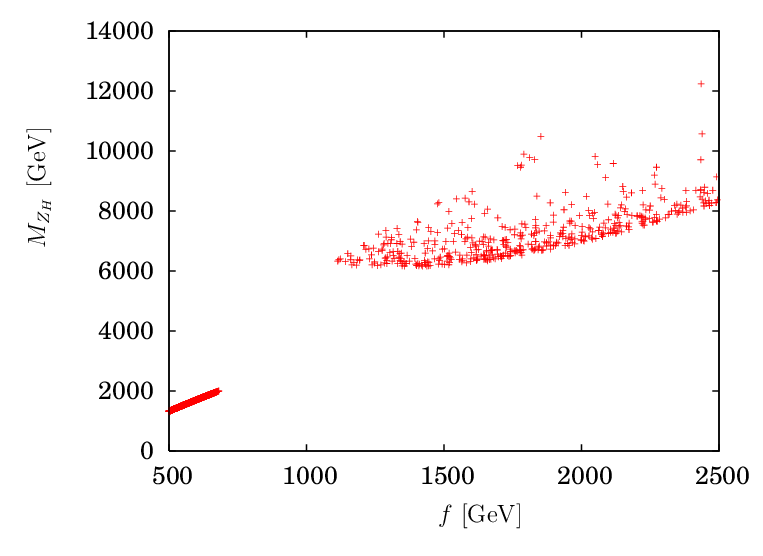}}
\subfigure[]{\includegraphics[width=2 in]{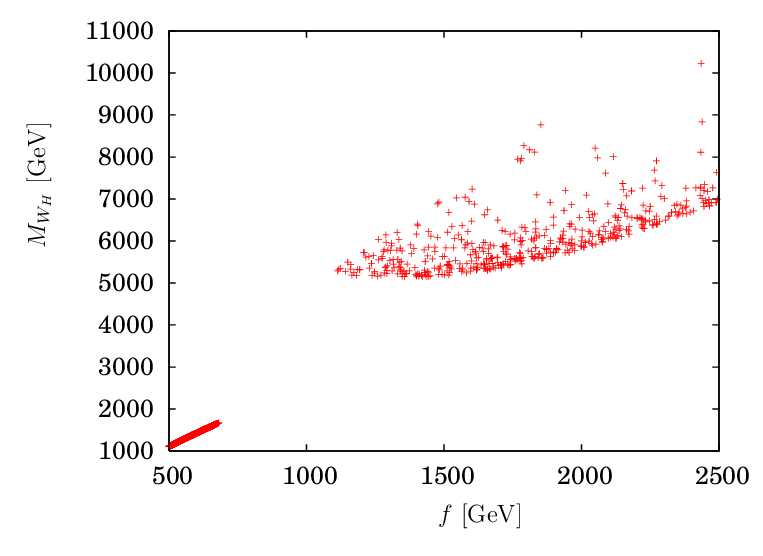}} 
\caption{\label{fig:topzw}
(a) The scatter plot for the  heavy top mass
with the horizontal axis being the scale parameter $f$. 
(b) The scatter plot for the heavy $Z$ boson mass
with the horizontal axis being the scale parameter $f$.
(c)  The scatter plot for the heavy $W$-boson mass
with the horizontal axis being the scale parameter $f$.}
\end{figure}

This region of parameter space can provide constraints on the
masses of many particles appear in this model. First, let us
consider the masses of the heavy top and heavy gauge bosons. As
shown in Figure~\ref{fig:topzw}, their masses generically increase
as $f$ increases. The mass of the heavy top quark is uniquely
determined when $f, \hat f$ and $M$ are fixed. So does the top Yukawa
coupling. Basically $\hat f$ is determined by the electroweak
symmetry breaking condition, but  their $M$ and $\mu_r$ dependence
provoke the ambiguity on its value. For small $f$ region, since
$M$ is also very small, the $M_T$ is almost determined by $f$
alone. It appears as straight line in Figure~\ref{fig:topzw}\,(a).
For large $f$ region, it becomes spread due to the top mixing
angles. The plots of the heavy $Z$ and $W$ boson masses versus $f$
are quite similar to that of the heavy top mass versus $f$. In the
case of heavy $W$ boson, the strongest constraint come from $K_L -
K_S$ mixing. The strongest bound ever known is  $m_{W_H} > 1.6$
TeV, with the assumption of $g_L = g_R$.~\cite{Beall:1981ze} This
can exclude some region from Figure~\ref{fig:topzw}(c). In this
case, small $f$ region can be completely excluded, but the
analysis of Ref.~\cite{Beall:1981ze} did not include the higher
order QCD corrections and used vacuum insertion to obtain the
matrix element. So we will not consider that bound seriously here.
Detailed study including QCD corrections and others is being done
by authors of Ref.~\cite{Goh:2006wj}. If the lower bound for $f$
is confirmed, we can give the lower bound for $f$ as $1.1 ~{\rm
TeV}$ from our calculation of the $\rho$-parameter and also for
many particles appear in the model. Another constraints on the
$m_{W_H}$ from CDF and D0 are about $650 \sim 786 {\rm GeV}$, as
lower bound~\cite{Affolder:2001gr,Abachi:1995yi}. For Our results
remain safe from these experimental bounds. Heavy $Z$ boson has
also been studied in detail by many experimentalists. Current
experimental bound is about $500 \sim 800 {\rm GeV}$ from
precision measurements~\cite{Yao:2006px} and $\sim 630 {\rm GeV}$
from CDF~\cite{Yao:2006px}. In this case, also safe is the mass of
heavy $Z$ boson.

\begin{figure}[htb!]
\centering
\subfigure[]{
\includegraphics[width=2 in]{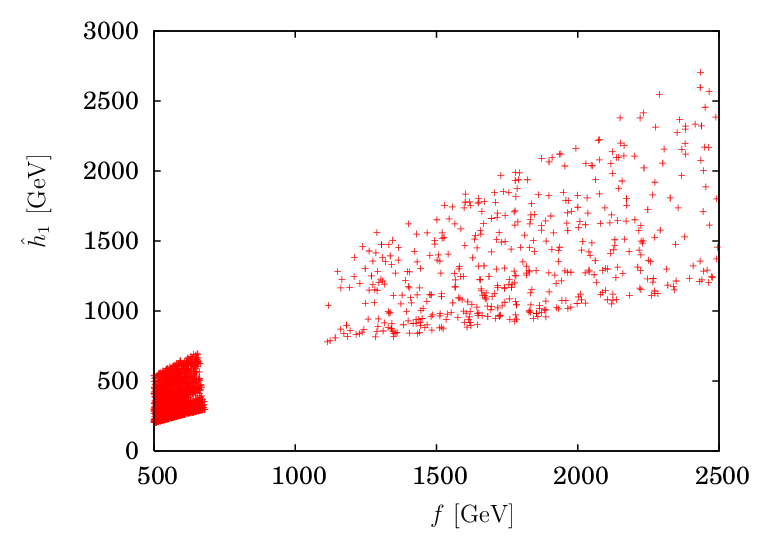}}
\subfigure[]{
\includegraphics[width=2 in]{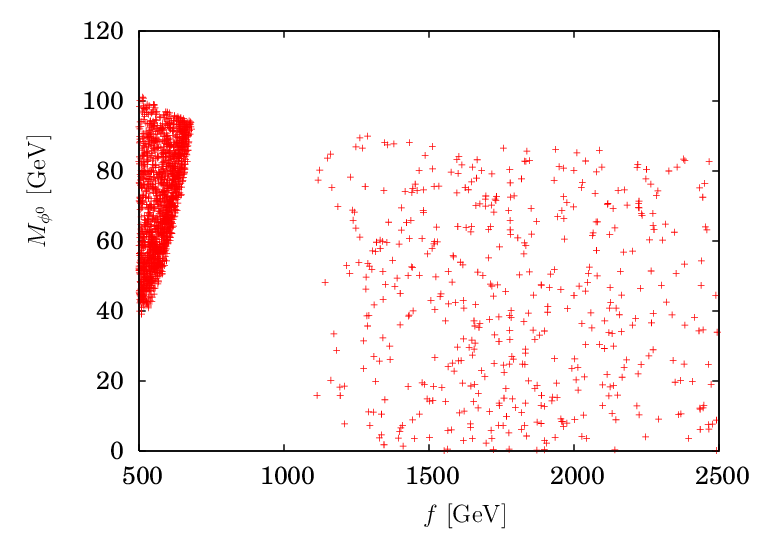}}
\subfigure[]{
\includegraphics[width=2 in]{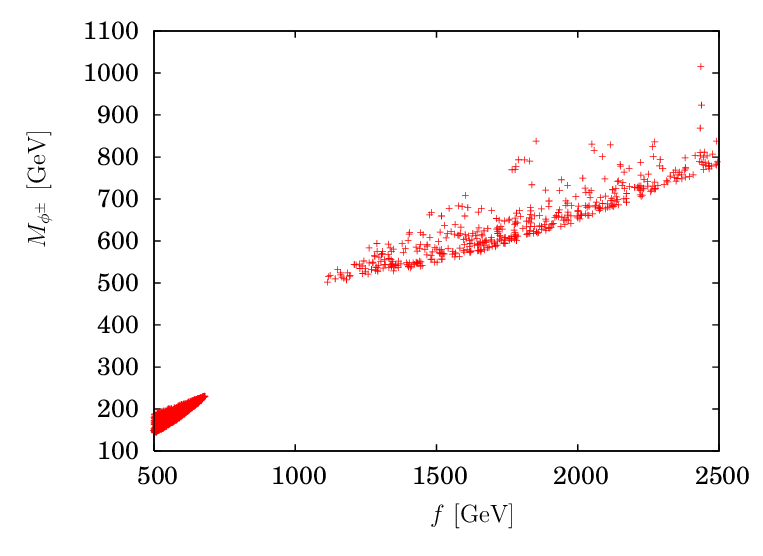}} 
\caption{\label{fig:hphiphi}
(a) The scatter plot for the mass of $\hat h_1$
with the horizontal axis being the scale parameter $f$.
It is similar to that of $\hat h_2$. 
(b) The scatter plot for the mass of $\phi^0$
with the horizontal axis being the scale parameter $f$.
(c)  The scatter plot for the mass of $\phi^+1$
with the horizontal axis being the scale parameter $f$.}
\end{figure}

\begin{figure}[htb!]
\includegraphics[width=4 in]{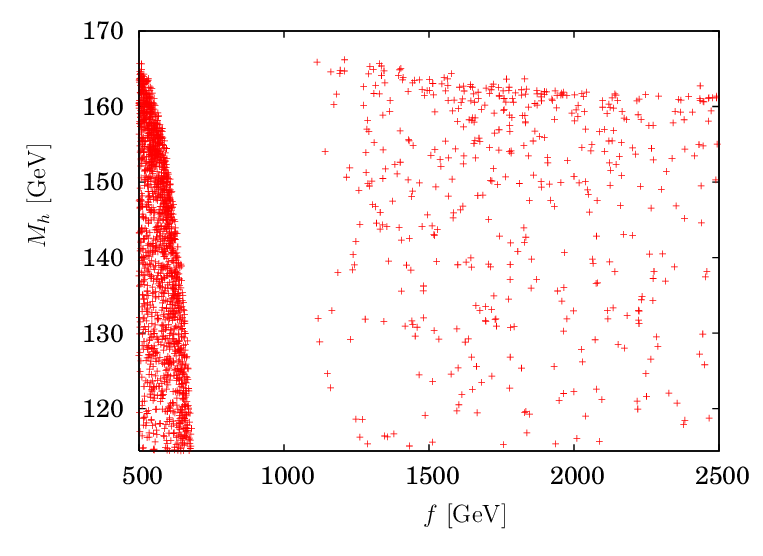}
\caption{\label{fig:higgs}The scatter plot for the SM Higgs mass
with the horizontal axis being the scale parameter $f$.}
\end{figure}

With the parameters allowed by the $\rho$-parameter, the masses of
new scalar bosons $\hat h_{1,2}, \phi^0$ and $ \phi^\pm$ can be
constrained. $\hat h_{1,2}$ have almost degenerate masses, and are
dependent on both $\mu_r$ and $\hat \mu$, unlike the
$\phi^{0,\pm}$ which depend only on $\mu_r$.
Their masses are substantially constrained according to the value of $f$.
Unfortunately, we cannot give a lower bound on the mass of $\phi^0$.
In fact, its mass, though it is quite small, arise from radiative corrections.
For $\phi^\pm$, the loop effects are rather large so they are heavier than $\phi^0$
as shown in Figure~\ref{fig:hphiphi}.
The distribution of the SM Higgs mass as a function of $f$ is shown in Figure~\ref{fig:higgs}.
As for the lower bound of the SM Higgs mass  we adopt the LEP bound for
Higgs mass, $114.4~{\rm GeV}$~\cite{Barate:2003sz}, since its
structure is same as the SM.
Its upper bound is approximately given as $167 {\rm GeV}$.

We summarize the results of our analysis as follows. With the
observed $\rho$-parameter, the allowed parameter space is divided into two separate
regions: $f$ smaller than about 670 GeV and larger than about 1.1 TeV.
We can give the mass bounds of the particles in the LRTH model for either region.
But the heavy gauge bosons remain safe from the experimental constraints.
Unlike the other particles, we cannot set a lower bound for the neutral $\phi^0$ scalar.
The loop corrections play an important role on the charged $\phi^\pm$ scalars,
yielding mass difference between the charged and neutral scalars.
Further analysis is required to reduce the allowed region.
If the small $f$ region is excluded, for example by Ref.~\cite{Goh:2006wj},
we can provide exact lower bounds for the masses of $T, Z_H, W_H, \hat h_{1,2}$, and $ \phi^\pm$.
But even in that case, we cannot do so for $\phi^0$ and SM Higgs boson.
\section{Conclusion}
The left right twin Higgs model is a concrete realization of the
twin Higgs mechanism. The model predicts a heavy top quark, heavy
gauge and various scalar bosons along with a light SM Higgs boson, and
will in turn yield rich phenomenology of the new particles at the
LHC. We have performed an indirect search for the existence of the
particles. The heavy Top and new scalars contribute significantly
to the isospin violating the $\rho$-parameter. One-loop radiative corrections
to the $\rho$-parameter reduce parameter space of the model and
can set rough bounds for the masses of the heavy particles. In particular,
we demonstrated that the symmetry breaking parameter $f$ can be either
smaller than $660~{\rm GeV}$ or larger than $1.1~{\rm TeV}$,
which is a crucial region in parameter space.
More analysis on other one-loop processes as well as study of collider physics
is mandatory to further reduce the region of parameter space.

\begin{acknowledgments}
Both J.Y.~Lee and D.W.~Jung are supported in part by NRF Research Grant 2012R1A2A1A01006053.
J.Y.~Lee is supported in part by Basic Science Research Program through the National Research Foundation of Korea(NRF)
funded by the Ministry of Education, Science and Technology(2011-0003974).
\end{acknowledgments}

\appendix
\section{Coupling constants of the LRTH model}
We summarize the relevant coupling constants relevant to our calculation.
The gauge-fermion interaction is given by
\begin{align}
{\mathcal L}&=i\bar\psi_1\gamma_\mu(g_V+g_A\gamma_5)\psi_2X^\mu\nonumber\\
&=i\bar\psi_1\gamma_\mu(c_LP_L+c_RP_R)\psi_2X^\mu,
\end{align}
where $P_{L,R}=\frac{1}{2}(1\mp\gamma_5)$ are the projection operators.
We make a list of the gauge coupling constants of the fermions in Table~\ref{table1}.
\begin{table}
\caption{Relevant coupling constants $X\bar \psi\psi$.
The mixing angles $C_L=\cos\alpha_L, C_R=\cos\alpha_R$, etc are given in Eq.~(\ref{eq:mixtop}).}
\begin{ruledtabular}
\begin{tabular}{|c|c|c|}\hline
$X\bar \psi\psi$ &  & \\ \hline
$W\bar tb$ & $c_L=eC_L/(\sqrt{2}s_w)$ & $c_R=0$ \\
$W\bar Tb$ & $c_L=eS_L/(\sqrt{2}s_w)$ & $c_R=0$\\\hline
$Z\bar tt$ & $g_V=e(\frac{1}{4}C^2_L-\frac{2}{3}s_w^2)/(c_ws_w)$
 & $g_A=-\frac{1}{4}eC_L^2/(c_ws_w)$ \\
$Z\bar bb$ & $g_V=e(-\frac{1}{2}+\frac{2}{3}s^2_w)/(2c_ws_w)$ & $g_A=e/(4c_ws_w)$ \\
$Z\bar TT$ & $g_V=e(\frac{1}{4}S^2_L-\frac{2}{3}s_w^2)/(c_ws_w)$
 & $g_A=-\frac{1}{4}eS_L^2/(c_ws_w)$ \\
$Z\bar Tt$ & $c_L=eC_LS_L/(2c_ws_w)$ &
$c_R=ef^2x^2s_wC_RS_R/(2\hat f^2c^3_w)$\\\hline $A\bar ff$ &
$g_V=eQ_f$ & $g_A=0$ \\ \hline
\end{tabular}\label{table1}
\end{ruledtabular}
\end{table}
The other gauge-scalar interactions are also taken into account.
We choose the unitary gauge where all gauge-scalar mixing terms
vanish. The various gauge coupling constants of the scalar fields
are given in Table~\ref{table2},~\ref{table3}, and~\ref{table4}.
\begin{table}
\caption{Relevant gauge coupling of the scalar fields, $C_{XSS}$.
$p1$, $p2$ and $p3$ refer to the incoming momentum  of the first, second and third particle, respectively~\cite{Han:2003wu}.}
\begin{ruledtabular}
\begin{tabular}{|c|c||c|c|}\hline $XSS$ & $C_{XSS}$& $XSS$
&$C_{XSS}$
\\\hline $W^+\hat h^\dagger_1\hat h_2$ &
$-e(p_2-p_3)_\mu/(\sqrt{2}s_w)$
& $A\hat h^\dagger_1\hat h_1$ & $-e(p_2-p_3)_\mu$ \\
$Z\hat h^\dagger_1\hat h_1$ & $-e(c^2_w-s^2_w)(p_2-p_3)_\mu/(2c_ws_w)$
&  $Z\hat h^\dagger_2\hat h_2$ & $e(p_2-p_3)_\mu/(2c_ws_w)$\\
$Z\phi^-\phi^+$ & $e(p_2-p_3)_\mu s_w/c_w$ & $A\phi^-\phi^+$ & $-e(p_2-p_3)_\mu$ \\
$Zh\phi^0$ & $iex p_{1\mu}/(6c_ws_w)$  & & \\\hline
\end{tabular}\label{table2}
\end{ruledtabular}
\end{table}
\begin{table}
\caption{Relevant gauge coupling constants of the scalar fields,
$C_{XXSS}$~\cite{Han:2003wu}.}
\begin{ruledtabular}
\begin{tabular}{|c|c||c|c|}\hline $XXSS$ & $C_{XXSS}$& $XXSS$ &
$C_{XXSS}$\\\hline
$W^+W^-hh$ & $e^2/(2s^2_w)$ & $ZZhh$ & $e^2/(2c^2_ws^2_2)$ \\
$W^+W^-\phi^0\phi^0$ & $-e^2x^2/(54s^2_w)$ & $ZZ\phi^0\phi^0$ & $-e^2x^2/(54c^2_ws^2_w)$\\
$W^+W^-\phi^+\phi^-$ & $-e^2x^2/(6s^2_w)$ & $ZZ\phi^+\phi^-$ & $2e^2s^2_w/c^2_w$\\
$W^+W^-\hat h_1^\dagger\hat h_1$ & $e^2/(2s^2_w)$ & $ZZ\hat h_1^\dagger\hat h_1$ & $e^2c2^4_w/(2c^2_ws^2_w)$\\
$W^+W^-\hat h_2^\dagger\hat h_2$ & $e^2/(2s^2_w)$ & $ZZ\hat h_2^\dagger\hat h_2$ & $e^2/(2c^2_ws^2_w)$\\\hline
$AA\hat h_1^\dagger\hat h_1$ & $2e^2$ & $AA\phi^+\phi^-$ & $2e^2$\\\hline
$ZA\phi^+\phi^-$ & $-2e^2s_w/c_w$ & $ZA\hat h_1^\dagger\hat h_1$ & $e^2c2^2_w/(c_ws_w)$ \\
$ZW^+\hat h^\dagger_1\hat h_2$ & $-e^2/(\sqrt{2}c_w)$ & $ZW^+\hat
h_1^\dagger \hat h_2$ & $e^2/(\sqrt{2}s_w)$\\ \hline
\end{tabular}\label{table3}
\end{ruledtabular}
\end{table}
\begin{table}
\caption{Relevant gauge coupling constants of the scalar fields,
$C_{XXS}$~\cite{Han:2003wu}.}
\begin{ruledtabular}
\begin{tabular}[c]{|c|c||c|c|} \hline $X_1X_2S$ & $C_{X_1X_2S}$&
$X_1X_2S$ &$C_{X_1X_2S}$ \\\hline
$W^+W^-h$ & $eM_W/s_w$ & $ZZh$ & $eM_W/(c^2_ws_w)$ \\
$ZZ_Hh$ & $ e^2fx/(\sqrt{2}c^2_wc2_w)$ & & \\\hline
\end{tabular}\label{table4}
\end{ruledtabular}
\end{table}
\section{One-loop integrals}
We list scalar integrals relevant for one-loop
Feynman diagrams. The one-loop scalar integrals are decomposed in
terms of Passarino-Veltman functions ~\cite{Passarino:1978jh}
which are defined in $d=4-2\epsilon$ dimensions,
\begin{align}
Q^{4-d}\int \frac{d^dk}{(2\pi)^d}\frac{1}{k^2-m^2+i\epsilon}&\equiv \frac{i}{16\pi^2}A_0(m^2),\\
Q^{4-d}\int \frac{d^dk}{(2\pi)^d}\frac{1}{(k^2-m^2_1+i\epsilon)((k-p)^2-m^2_2+i\epsilon)}
&\equiv \frac{i}{16\pi^2}B_0(p^2,m^2_1,m^2_2),\\
Q^{4-d}\int \frac{d^dk}{(2\pi)^d}\frac{k_\mu}{(k^2-m^2_1+i\epsilon)((k-p)^2-m^2_2+i\epsilon)}
&\equiv \frac{i}{16\pi^2}p_\mu B_1(p^2,m^2_1,m^2_2),\\
Q^{4-d}\int \frac{d^dk}{(2\pi)^d}\frac{k_\mu k_\nu}{(k^2-m^2_1+i\epsilon)((k-p)^2-m^2_2+i\epsilon)}
&\equiv \frac{i}{16\pi^2}[g_{\mu\nu} B_{22}(p^2,m^2_1,m^2_2)\nonumber\\
&\qquad \quad +\,p_\mu p_\nu  B_{11}(p^2,m^2_1,m^2_2)],
\end{align}
where $Q$ is the renormalization scale and
$1/\hat\epsilon=(4\pi)^\epsilon\Gamma(1+\epsilon)/\epsilon$.
We also define the following integrals,
\begin{align}
I_1(a)&\equiv\int^1_0dx\,\ln[1-ax(1-x)],\\
I_3(a)&\equiv\int^1_0dx\,x(1-x)\ln[1-ax(1-x)],\\
I_4(a,b)&\equiv\int^1_0dx\,\ln[1-x+ax-bx(1-x)],\\
I_5(a,b)&\equiv\int^1_0dx\,x\ln[1-x+ax-bx(1-x)],\\
I_6(a,b)&\equiv\int^1_0dx\,(1-x)\ln[1-x+ax-bx(1-x)],\\
I_7(a,b)&\equiv\int^1_0dx\,x(1-x)\ln[1-x+ax-bx(1-x)].
\end{align}

\section{Gauge boson self-energies in the LRTH model}
We calculate the four gauge boson self-energies,
$\Pi^{\gamma\gamma'}(0)$, $\Pi^{\gamma Z}(M^2_Z)$, $\Pi^{WW}(0)$
and $\Pi^{ZZ}(M^2_Z)$. In general, the gauge independence in the
bosonic sector can be retained by using the pinch technique or by
using the background field method
\cite{Degrassi:1992ue,Denner:1994nn}. In our calculations, there
are three one-loop diagrams involved with an internal gauge boson
propagator. In these diagrams we take only gauge invariant parts
which are proportional to ln$(M^2_S)/16\pi^2$.

\subsection{Contributions to $\Pi^{\gamma\gamma'}(0)$}
\begin{figure}[htb!]
\includegraphics[width=6 in]{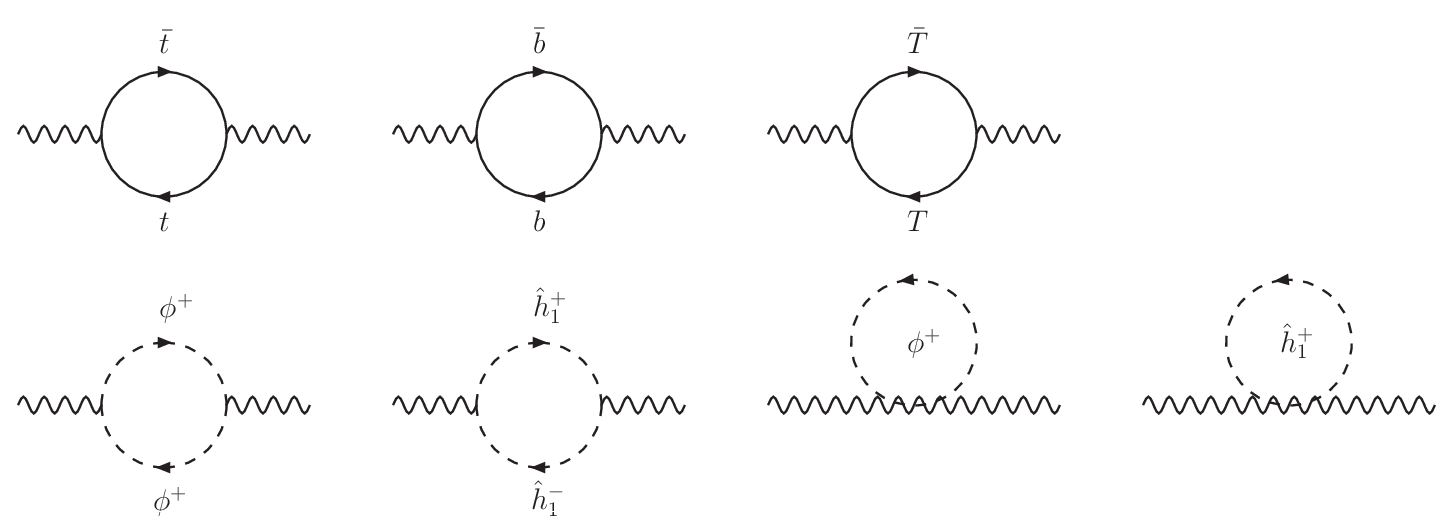}
\caption{The one-loop corrections to the self-energy $\Pi^{\gamma\gamma}$.}
\label{fig:gg}
\end{figure}

The one-loop corrections to the self-energy $\Pi^{\gamma\gamma}$ of the LRTH model
are shown in Figure~\ref{fig:gg}.
The total contribution to the self-energy is
\begin{equation}
\Pi^{\gamma\gamma'}(0)=\frac{\alpha}{4\pi}\Big[\frac{16}{9}\ln\frac{Q^2}{m^2_t}
+\frac{4}{9}\ln\frac{Q^2}{m^2_b}+\frac{16}{9}\ln\frac{Q^2}{m^2_T}
+\frac{1}{3}\ln\frac{Q^2}{m^2_{\phi^+}}
+\frac{1}{3}\ln\frac{Q^2}{m^2_{\hat h_1}}
+\frac{14}{3\hat\epsilon}\Big].
\end{equation}

\subsection{Contributions to $\Pi^{\gamma Z}(M^2_Z)$}
\begin{figure}[htb!]
\includegraphics[width=6 in]{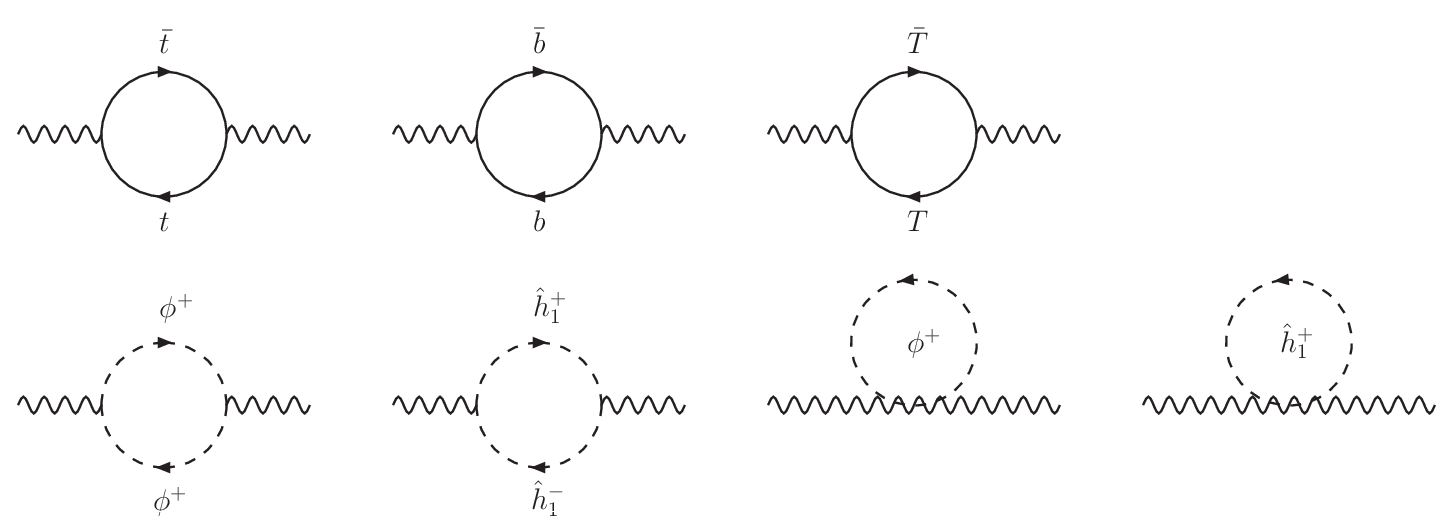}
\caption{The one-loop corrections to the self-energy $\Pi^{\gamma Z}$.}
\label{fig:gz}
\end{figure}

The one-loop corrections to the self-energy $\Pi^{\gamma Z}(M^2_Z)$ are shown in Figure~\ref{fig:gz}.
These are (i) fermionic loops having $(\bar tt),(\bar TT)$ and $(\bar bb)$,
(ii) the scalar loops due to $SSV$ coupling, $(\phi^+\phi^-)$, $(\hat h^+_1\hat h^-_1)$, 
and (iii) the $\phi^+$ and $\hat h^+_1$ scalar loops due to $SSVV$ quartic couplings.
The contributions to $\Pi^{\gamma Z}(M^2_Z)$ due to the fermion loops through the couplings
in Table~\ref{table1} are
\begin{align}
\Pi^{\gamma Z}_{\bar t t}(M^2_Z)
&=\frac{N_c\alpha}{\pi}\frac{2}{3s_wc_w}\Big(\frac{1}{4}C^2_L-\frac{2}{3}s^2_w\Big)
M^2_Z\Big[\frac{1}{3}\Big(\ln\frac{Q^2}{m^2_t}+\frac{1}{\hat \epsilon}\Big)
-2I_3\Big(\frac{M^2_Z}{m^2_t}\Big)\Big],\\
\Pi^{\gamma Z}_{\bar T T}(M^2_Z)
&=\frac{N_c\alpha}{\pi}\frac{2}{3s_wc_w}\Big(\frac{1}{4}S^2_L-\frac{2}{3}s^2_w\Big)
M^2_Z\Big[\frac{1}{3}\Big(\ln\frac{Q^2}{m^2_T}+\frac{1}{\hat \epsilon}\Big)
-2I_3\Big(\frac{M^2_Z}{m^2_T}\Big)\Big],\\
\Pi^{\gamma Z}_{\bar bb}(M^2_Z)
&=\frac{N_c\alpha}{4\pi}\frac{1}{3s_wc_w}\Big(\frac{1}{2}-\frac{2}{3}s^2_w\Big)
M^2_Z\Big[\frac{1}{3}\Big(\ln\frac{Q^2}{m^2_b}+\frac{1}{\hat \epsilon}\Big)
-2I_3\Big(\frac{M^2_Z}{m^2_b}\Big)\Big].
\end{align}

The contributions to $\Pi^{\gamma Z}(M^2_Z)$ from the scalar loops
through the couplings in Table~\ref{table2}  are
\begin{align}
\Pi^{\gamma Z}_{\hat h_1 \hat h_1}(M^2_Z)
&=-\frac{\alpha}{4\pi}\frac{c2^2_w}{c_ws_w}\Big[\Big(M^2_{\hat h_1}-\frac{1}{6}M^2_Z\Big)
\Big(\ln\frac{Q^2}{M^2_{\hat h_1}}+\frac{1}{\hat \epsilon}\Big)\nonumber\\
&\qquad+\Big(\frac{1}{6}M^2_Z-\frac{2}{3}M^2_{\hat h_1}\Big) I_1\Big(\frac{M^2_Z}{M^2_{\hat h_1}}\Big)
+M^2_{\hat h_1}-\frac{1}{9}M^2_Z\Big],\label{eq:gzh1h1}\\
\Pi^{\gamma Z}_{\phi^+\phi^-}(M^2_Z)
&=\frac{\alpha}{2\pi}\frac{s_w}{c_w}\Big[\Big(M^2_{\phi^+}-\frac{1}{6}M^2_Z\Big)
\Big(\ln\frac{Q^2}{M^2_{\phi^+}}+\frac{1}{\hat \epsilon}\Big)\nonumber\\
&\qquad+\Big(\frac{1}{6}M^2_Z-\frac{2}{3}M^2_{\phi^+}\Big) I_1\Big(\frac{M^2_Z}{M^2_{\phi^+}}\Big)
+M^2_{\phi^+}-\frac{1}{9}M^2_Z\Big].\label{eq:gzpppm}
\end{align}

The contributions to $\Pi^{\gamma Z}(M^2_Z)$ from the scalar loops
through the couplings in Table~\ref{table3} are
\begin{align}
\Pi^{\gamma Z}_{\hat h_1 \hat h_1}(M^2_Z)
&=\frac{\alpha}{4\pi}\frac{c2^2_w}{c_ws_w}\Big[1+\ln\frac{Q^2}{M^2_{\hat h_1}}
+\frac{1}{\hat \epsilon}\Big]M^2_{\hat h_1},\label{eq:gzh1h12}\\
\Pi^{\gamma Z}_{\phi^+\phi^-}(M^2_Z)
&=-\frac{\alpha}{2\pi}\frac{s_w}{c_w}\Big[1+\ln\frac{Q^2}{M^2_{\phi^+}}
+\frac{1}{\hat \epsilon}\Big]M^2_{\phi^+}.\label{eq:gzpppm2}
\end{align}

The terms proportional to $M^2_{\hat h_1}$ and $M^2_{\hat h_1}\ln(Q^2/M^2_{\hat h_1})$
in Eq.~(\ref{eq:gzh1h1}) and (\ref{eq:gzh1h12}) cancel between themselves and so do the
terms proportional to $M^2_{\phi^+}$ and $M^2_{\phi^+}\ln(Q^2/M^2_{\phi^+})$
in Eq.~(\ref{eq:gzpppm}) and (\ref{eq:gzpppm2}).
For $p^2=0$, it can be easily checked that the total fermionic and scalar contributions
vanish individually. As expected in the unitary gauge no mixing between the two gauge bosons
takes place at one-loop due to 
\begin{equation}
\Pi^{\gamma Z}(0)=0.
\end{equation}

\subsection{Contributions to $\Pi^{WW}(0)$}
The contributions to $\Pi^{WW}(0)$ from the fermion loops through the couplings
in Table~\ref{table1} are given as follows,
\begin{align}
\Pi^{WW}_{\bar t b}(0)
&= \frac{N_c\alpha}{4\pi}\frac{C^2_L}{2s^2_w} f_1(m^2_t,m^2_b),\\
\Pi^{WW}_{\bar T b}(0)
&= \frac{N_c\alpha}{4\pi}\frac{S^2_L}{2s^2_w} f_1(m^2_T,m^2_b),
\end{align}
where $1/\hat \epsilon$ terms are omitted, and $f_1(m^2_1,m^2_2)$ is defined as
\begin{equation}
f_1(m^2_1,m^2_2)=\frac{1}{2}(m^2_1+m^2_2)+\frac{m^4_1}{m^2_1-m^2_2}\ln\frac{Q^2}{m^2_1}
-\frac{m^4_2}{m^2_1-m^2_2}\ln\frac{Q^2}{m^2_2}.
\end{equation}
The contributions to $\Pi^{WW}(0)$ from the scalar loops through the couplings
in Table~\ref{table3} are given as,
\begin{align}
\Pi^{WW}_h(0)
&=\frac{\alpha}{16\pi}\frac{1}{s^2_w}\big[1+\ln\frac{Q^2}{M^2_h}+\frac{1}{\hat \epsilon}\big]M^2_h,\\
\Pi^{WW}_{\hat h_1^+}(0)
&=\frac{\alpha}{8\pi}\frac{1}{s^2_w}\big[1+\ln\frac{Q^2}{M^2_{\hat h_1}}+\frac{1}{\hat \epsilon}\big]M^2_{\hat h_1},\label{eq:wwh1h1}\\
\Pi^{WW}_{\hat h_2^0}(0)
&=\frac{\alpha}{16\pi}\frac{1}{s^2_w}\big[1+\ln\frac{Q^2}{M^2_{\hat h_2}}+\frac{1}{\hat \epsilon}\big]M^2_{\hat h_2},\label{eq:wwh2h2}\\
\Pi^{WW}_{\phi^+}(0)
&=-\frac{\alpha}{24\pi}\frac{x^2}{s^2_w}\big[1+\ln\frac{Q^2}{M^2_{\phi^+}}+\frac{1}{\hat \epsilon}\big]M^2_{\phi^+},\label{eq:wwpp}\\
\Pi^{WW}_{\phi^0}(0)
&=-\frac{\alpha}{432\pi}\frac{x^2}{s^2_w}\big[1+\ln\frac{Q^2}{M^2_{\phi^0}}+\frac{1}{\hat \epsilon}\big]M^2_{\phi^0}.\label{eq:wwp0}
\end{align}
Note that $\Pi^{WW}_{\phi^+}$ and $\Pi^{WW}_{\phi^0}$ are much smaller than the other contributions
due to the suppression factor $x^2$.
\begin{figure}[htb!]
\includegraphics[width=6 in]{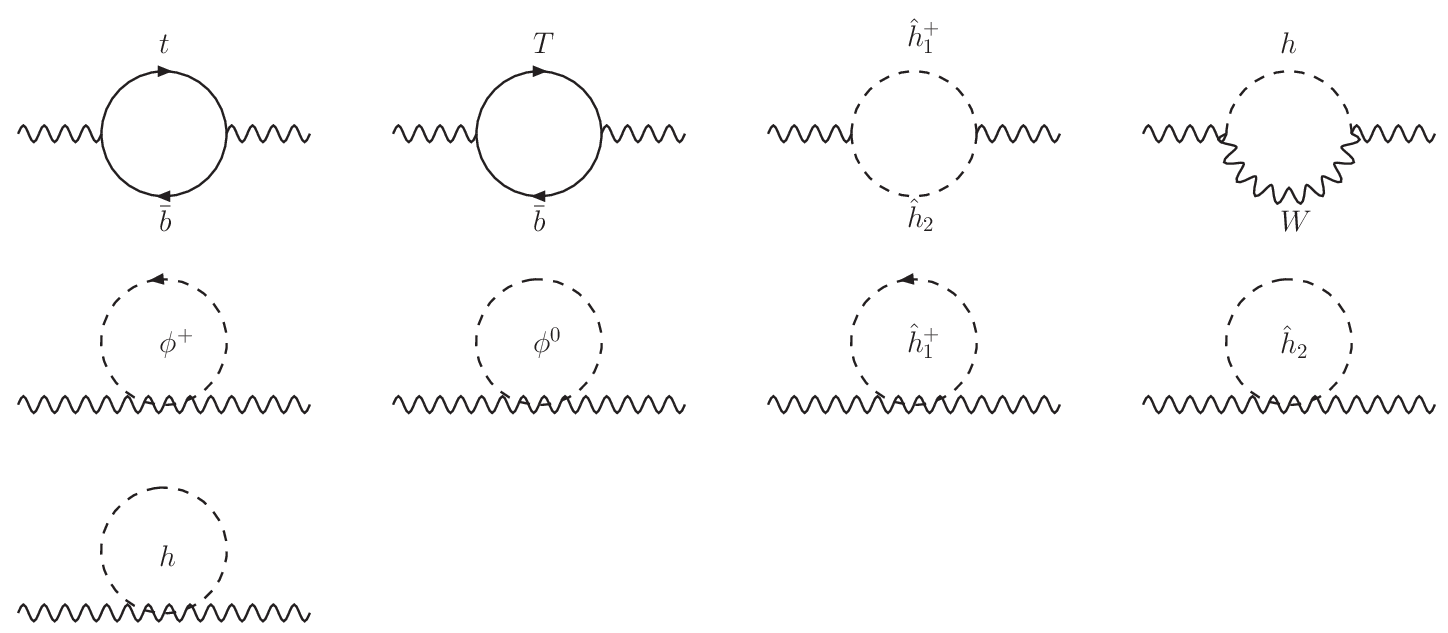}
\caption{The one-loop corrections to the self-energy $\Pi^{WW}$.}
\label{fig:ww}
\end{figure}

The contribution to $\Pi^{WW}(0)$ from the scalar loops through the couplings
in Table~\ref{table2} has the following form
\begin{equation}\label{eq:wwh1h2}
\Pi^{WW}_{\hat h_1\hat h_2}(0)=-\frac{\alpha}{2\pi}\frac{1}{s^2_w}g_1(M^2_{\hat h_1},M^2_{\hat h_2}),
\end{equation}
where $g_1(m^2_1,m^2_2)$ is defined as
\begin{equation}
g_1(m^2_1,m^2_2)=\frac{3}{8}(m^2_1+m^2_2)+\frac{1}{4(m^2_1-m^2_2)}\big[m^4_1\ln\frac{Q^2}{m^2_1}-m^4_2\ln\frac{Q^2}{m^2_2}\Big].
\end{equation}
The terms proportional to $M^2_{\hat h_1}$ and $M^2_{\hat h_1}\ln(Q^2/M^2_{\hat h_1})$
in Eq.~(\ref{eq:wwh1h1}) and (\ref{eq:wwh1h2}) cancel partially between themselves and so do the
terms proportional to $M^2_{\hat h_2}$ and $M^2_{\hat h_2}\ln(Q^2/M^2_{\hat h_2})$
in Eq.~(\ref{eq:wwh2h2}) and (\ref{eq:wwh1h2}).
Although the terms proportional to $M^2_{\phi^{+,0}}$ and $M^2_{\phi^{+,0}}\ln(Q^2/M^2_{\phi^{+,0}})$
in Eq.~(\ref{eq:wwpp}) and (\ref{eq:wwp0}) do not cancel out, 
their coefficients are significantly small and so are their contributions to $\Pi^{WW}(0)$.

The contribution to $\Pi^{WW}(0)$ from the SM Higgs-W boson loops has the following form
\begin{align}
\Pi^{WW}_{hW}(0)
&=\frac{\alpha}{4\pi}\frac{M^2_W}{s^2_w}\big[\frac{5}{8}-\frac{3}{8}\frac{M^2_h}{M^2_W}
+\frac{3}{4}\frac{M^2_h}{M^2_W-M^2_h}\ln\frac{Q^2}{M^2_W}\nonumber\\
&+\frac{M^2_h}{M^2_W-M^2_h}(-1+\frac{M^2_h}{M^2_W})\ln\frac{Q^2}{M^2_h}
+(1-\frac{M^2_W+M^2_h}{4M^2_W})\frac{1}{\hat\epsilon}\big].
\end{align}
We take only the contribution proportional to $\ln(M^2_S)/16\pi^2$, which is gauge invariant,
\begin{equation}
\Pi^{WW}_{hW}(0)=\frac{\alpha}{4\pi}\frac{M^2_W}{s^2_w}\ln\frac{Q^2}{M^2_h}
\Big[\frac{M^2_h}{M^2_W-M^2_h}(-1+\frac{M^2_h}{M^2_W})\Big].
\end{equation}
\subsection{Contributions to $\Pi^{ZZ}(M^2_Z)$}
The one-loop corrections to the self-energy function $\Pi^{ZZ}(p^2)$ are
shown in Figure~\ref{fig:zz}.
The complete list of fermionic contributions to the self-energy function are given below.
\begin{align}
\Pi^{ZZ}_{(\bar Tt)}(M^2_Z)&=
\frac{N_c\alpha}{4\pi}\frac{1}{ c^2_w} \Big[\frac{C_L^2S_L^2}{s^2_w}+\frac{C_R^2 S_R^2 f^4 x^4}{\hat f^4 c^4_w}\Big]     \Big[\Big(\frac{1}{\hat \epsilon}+\ln\frac{Q^2}{m^2_t}\Big)\Big(\frac{M^2_Z}{6}-\frac{m^2_t+m^2_T}{4}\Big) \nonumber\\
&\quad -M^2_Z I_7\Big(\frac{m^2_T}{m^2_t},\frac{M^2_Z}{m^2_t}\Big)-\frac{m^2_T}{2} I_5 \Big(\frac{m^2_T}{m^2_t},\frac{M^2_Z}{m^2_t}\Big) -\frac{m^2_t}{2} I_6 \Big(\frac{m^2_T}{m^2_t},\frac{M^2_Z}{m^2_t}\Big) \Big]\nonumber\\
&\quad +\frac{N_c\alpha}{8\pi}\frac{C_LS_LC_RS_R}{c_w^4}\frac{x^2f^2}{\hat f^2}m_tm_T\Big[\frac{1}{\hat \epsilon}+\ln\frac{Q^2}{m^2_t}-I_4(\frac{m^2_T}{m^2_t},\frac{M^2_Z}{m^2_t})\Big],\nonumber\\
\Pi^{ZZ}_{(\bar tT)}(M^2_Z)&=\Pi^{ZZ}_{(\bar Tt)}(M^2_Z)(m_t\leftrightarrow m_T),\\
\Pi^{ZZ}_{(\bar tt)}(M^2_Z)&=
\frac{N_c\alpha}{\pi}\frac{1}{c^2_w s^2_w} \Big[\Big(\frac{1}{2}C^2_L-\frac{2}{3}s^2_w\Big)^2+\frac{4}{9}s^4_w\Big] \Big[\Big(\frac{1}{\hat \epsilon}+\ln\frac{Q^2}{m^2_t}\Big)\Big(\frac{M^2_Z}{6}-\frac{m^2_t}{2}\Big) \nonumber\\
&\quad -M^2_Z I_3\Big(\frac{M^2_Z}{m^2_t}\Big)-\frac{m^2_T}{2} I_1 \Big(\frac{M^2_Z}{m^2_t}\Big) \Big]\nonumber\\
& \quad - \frac{2N_c\alpha}{3\pi} \frac{1}{c^2_w}\Big(\frac{1}{2}C^2_L-\frac{2}{3}s^2_w\Big) m^2_t \Big[\frac{1}{\hat\epsilon}+\ln\frac{Q^2}{m^2_t}-
I_1\Big(\frac{M^2_Z}{m^2_t}\Big)\Big],\\
\Pi^{ZZ}_{(\bar TT)}(M^2_Z)&=
\frac{N_c\alpha}{\pi}\frac{1}{ c^2_w s^2_w} \Big[\Big(\frac{1}{2}S^2_L-\frac{2}{3}s^2_w\Big)^2+\frac{4}{9}s^4_w\Big] \Big[\Big(\frac{1}{\hat \epsilon}+\ln\frac{Q^2}{m^2_T}\Big)\Big(\frac{M^2_Z}{6}-\frac{m^2_T}{2}\Big) \nonumber\\
&\quad -M^2_Z I_3\Big(\frac{M^2_Z}{m^2_T}\Big)-\frac{m^2_T}{2} I_1 \Big(\frac{M^2_Z}{m^2_T}\Big) \Big]\nonumber\\
& \quad -\frac{2N_c\alpha}{3\pi} \frac{1}{c^2_w}\Big(\frac{1}{2}S^2_L-\frac{2}{3}s^2_w\Big) m^2_T \Big[\frac{1}{\hat\epsilon}+\ln\frac{Q^2}{m^2_T}-
I_1\Big(\frac{M^2_Z}{m^2_T}\Big)\Big],\\
\Pi^{ZZ}_{(\bar bb)}(M^2_Z)&=
\frac{N_c\alpha}{4\pi}\frac{1}{c^2_w s^2_w} \Big[\Big(-1+\frac{2}{3}s^2_w\Big)^2+\frac{4}{9}s^4_w\Big]  
\Big[\Big(\frac{1}{\hat \epsilon}+\ln\frac{Q^2}{m^2_b}\Big)\Big(\frac{M^2_Z}{6}-\frac{m^2_b}{2}\Big) \nonumber\\
&\quad -M^2_Z I_3\Big(\frac{M^2_Z}{m^2_b}\Big)-\frac{m^2_b}{2} I_1 \Big(\frac{M^2_Z}{m^2_b}\Big) \Big]\nonumber\\
&\quad +\frac{N_c\alpha}{6\pi} \frac{1}{c^2_w}\Big(-1+\frac{2}{3}s^2_w\Big) m^2_b \Big[\frac{1}{\hat\epsilon}+\ln\frac{Q^2}{m^2_b}-
I_1\Big(\frac{M^2_Z}{m^2_b}\Big)\Big].
\end{align}
The contributions to $\Pi^{ZZ}(M^2_Z)$ from the scalar loops through the couplings
in Table~\ref{table3} have the following form,
\begin{align}
\Pi^{ZZ}_{(h)}(M^2_Z)&=
\frac{\alpha}{16\pi}\frac{1}{c^2_ws^2_w}M^2_h
\Big[1+\ln\frac{Q^2}{M^2_h}+\frac{1}{\hat \epsilon}\Big],\\
\Pi^{ZZ}_{(\hat h_1)}(M^2_Z)&=
\frac{\alpha}{8\pi}\frac{c2^4_w}{c^2_ws^2_w}M^2_{\hat h_1}
\Big[1+\ln\frac{Q^2}{M^2_{\hat h_1}}+\frac{1}{\hat \epsilon}\Big],\label{eq:zzh1h1}\\
\Pi^{ZZ}_{(\hat h_2)}(M^2_Z)&=
\frac{\alpha}{8\pi}\frac{1}{c^2_ws^2_w}M^2_{\hat h_2}
\Big[1+\ln\frac{Q^2}{M^2_{\hat h_2}}+\frac{1}{\hat \epsilon}\Big],\label{eq:zzh2h2}\\
\Pi^{ZZ}_{(\phi^+)}(M^2_Z)&=
\frac{\alpha}{2\pi}\frac{s^2_w}{c^2_w}M^2_{\phi^+}
\Big[1+\ln\frac{Q^2}{M^2_{\phi^+}}+\frac{1}{\hat \epsilon}\Big],\label{eq:zzpp}\\
\Pi^{ZZ}_{(\phi^0)}(M^2_Z)&=
-\frac{\alpha}{8\pi}\frac{x^2}{54c^2_ws^2_w}M^2_{\phi^0}
\Big[1+\ln\frac{Q^2}{M^2_{\phi^0}}+\frac{1}{\hat \epsilon}\Big].
\end{align}

The contributions to $\Pi^{ZZ}(M^2_Z)$ from the scalar loops through the couplings
in Table~\ref{table2} have the following form,
\begin{align}
\Pi^{ZZ}_{(\hat h_1^\dagger \hat h_1)}(M^2_Z)&=
-\frac{\alpha}{8\pi}\frac{c2^4_w}{c^2_ws^2_w}\Big[\Big(M^2_{\hat h_1}-\frac{1}{6}M^2_Z\Big)
\Big(\ln\frac{Q^2}{M^2_{\hat h_1}}+\frac{1}{\hat \epsilon}\Big)\nonumber\\
&\quad +\Big(\frac{1}{6}M^2_Z-\frac{2}{3}M^2_{\hat h_1}\Big)I_1\Big(\frac{M^2_Z}{M^2_{\hat h_1}}\Big)
+M^2_{\hat h_1}-\frac{1}{9}M^2_Z\Big],\label{eq:zzh1h12}\\
\Pi^{ZZ}_{(\hat h_2^\dagger \hat h_2)}(M^2_Z)&=
-\frac{\alpha}{8\pi}\frac{1}{c^2_ws^2_w}\Big[\Big(M^2_{\hat h_2}-\frac{1}{6}M^2_Z\Big)
\Big(\ln\frac{Q^2}{M^2_{\hat h_2}}+\frac{1}{\hat \epsilon}\Big)\nonumber\\
&\quad +\Big(\frac{1}{6}M^2_Z-\frac{2}{3}M^2_{\hat h_2}\Big)I_1\Big(\frac{M^2_Z}{M^2_{\hat h_2}}\Big)
+M^2_{\hat h_2}-\frac{1}{9}M^2_Z\Big],\label{eq:zzh2h22}\\
\Pi^{ZZ}_{(\phi^+\phi^-)}(M^2_Z)&=
-\frac{\alpha}{2\pi}\frac{s^2_w}{c^2_w}\Big[\Big(M^2_{\phi^+}-\frac{1}{6}M^2_Z\Big)
\Big(\ln\frac{Q^2}{M^2_{\phi^+}}+\frac{1}{\hat \epsilon}\Big)\nonumber\\
&\quad +\Big(\frac{1}{6}M^2_Z-\frac{2}{3}M^2_{\phi^+}\Big)I_1\Big(\frac{M^2_Z}{M^2_{\phi^+}}\Big)
+M^2_{\phi^+}-\frac{1}{9}M^2_Z\Big],\label{eq:zzpp2}\\
\Pi^{ZZ}_{(h\phi^0)}(M^2_Z)&=0.
\end{align}
The terms proportional to $M^2_{\hat h_1}$ and $M^2_{\hat h_1}\ln(Q^2/M^2_{\hat h_1})$
in Eq.~(\ref{eq:zzh1h1}) and (\ref{eq:zzh1h12}) cancel between themselves and so do the
terms proportional to $M^2_{\hat h_2}$ and $M^2_{\hat h_2}\ln(Q^2/M^2_{\hat h_2})$
in Eq.~(\ref{eq:zzh2h2}) and (\ref{eq:zzh2h22}).
The terms proportional to $M^2_{\phi^+}$ and $M^2_{\phi^+}\ln(Q^2/M^2_{\phi^+})$
in Eq.~(\ref{eq:zzpp}) and (\ref{eq:zzpp2}) also cancel between themselves.

There are contributions of scalar-gauge boson loops to $\Pi^{ZZ}(M^2_Z)$. We take only the
contribution proportional to $\ln(M^2_S)/16\pi^2$, which is gauge invariant,
\begin{align}
\Pi^{ZZ}_{(Zh)}(M^2_Z)&=
\frac{\alpha}{8\pi}\frac{M^2_W}{c^4_ws^2_w}\ln\frac{Q^2}{M^2_h}
\Big[1-\frac{3M^2_h+2M^2_Z}{12M^2_Z}\Big],\\
\Pi^{ZZ}_{(Z_Hh)}(M^2_Z)&=
\frac{\alpha}{16\pi}\frac{f^2x^2}{c^4_wc2^2_w}\ln\frac{Q^2}{M^2_h}
\Big[1-\frac{3M^2_h+3M^2_{Z_H}-M^2_Z}{12M^2_{Z_H}}\Big].
\end{align}
\begin{figure}[htb!]
\includegraphics[width=6 in]{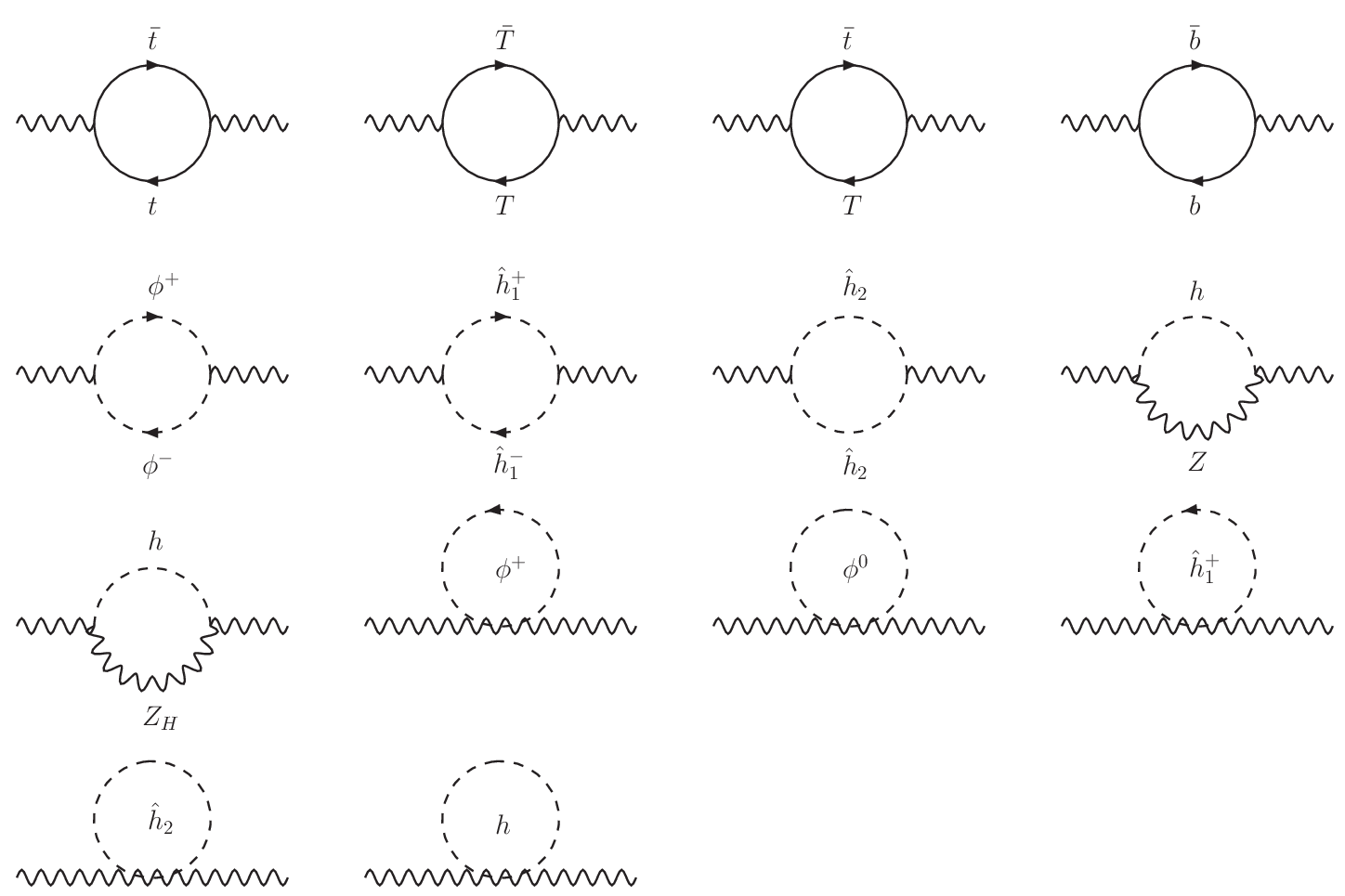}
\caption{The one-loop corrections to the self-energy function $\Pi^{ZZ}$.}
\label{fig:zz}
\end{figure}

\end{document}